\documentclass{iau_FM}
\usepackage{graphicx}
\usepackage{aas_macros}
\usepackage{physics}
\usepackage{subfigure}
\usepackage{amsmath}
\usepackage{wrapfig}
\usepackage{hyperref}
\usepackage{natbib}
\hypersetup{
    colorlinks=true,
    linkcolor=blue,
    filecolor=blue,
    urlcolor=blue,
    citecolor=blue,
   pdftitle={IAU FM9},
}

\title[FRBs to constrain PBHs comprised of dark matter] 
{Understanding constraints on primordial mass black holes made of dark matter using fast radio bursts}

\author[S. Kalita, S. Bhatporia \& A. Weltman]   
{Surajit Kalita$^1$,
 Shruti Bhatporia$^2$
 \and Amanda Weltman$^3$}

\affiliation{High Energy Physics, Cosmology \& Astrophysics Theory (HEPCAT) Group, \\ Department of Mathematics \& Applied Mathematics, \\ University of Cape Town, Cape Town 7700, South Africa \\ $^1$email: {\tt surajit.kalita@uct.ac.za} \\[\affilskip]
$^2$email: {\tt bhtshr001@myuct.ac.za}\\
$^3$email: {\tt amanda.weltman@uct.ac.za}}

\pubyear{2024}
\volume{FM9}  
\pagerange{1--3}
\jname{32nd IAU General Assembly}
\editors{A.C. Editor, B.D. Editor \& C.E. Editor, eds.}
\begin{document}

\maketitle

\begin{abstract}
In recent decades, a multitude of modified gravity theories have been proposed to address a variety of cosmological and astrophysical problems. While many of these theories remain viable, observational constraints on their parameters are increasingly stringent. Fast Radio Bursts (FRBs), in particular, have emerged as powerful probes of cosmology and fundamental physics. This study investigates the implications of a generic modified gravity theory for gravitational lensing by FRBs. By analyzing the dataset of CHIME/FRBs, we constrain the fraction of dark matter composed of primordial black holes within this theoretical framework. Furthermore, we demonstrate that modified gravity introduces a screening effect on gravitational lensing, analogous to the scattering effect of plasma on light rays.
\end{abstract}

\firstsection 
\section{Introduction}
Fast Radio Bursts (FRBs) can be used as a tool to understand different cosmological phenomena because of their distinct features, such as short pulse width, relatively high dispersion measure (DM), etc. While only one FRB-like pulse has been observed within our Galaxy \citep{2020Natur.587...54C}, the majority are extragalactic, many now with identified host galaxies. The progenitor mechanism is still uncertain, though we expect them to be associated with extreme astrophysical phenomena comprising compact objects \citep{2019PhR...821....1P}. In the future, multimessenger observations or gravitational wave and neutrino detection from FRB sites can help in identifying the progenitor mechanisms \citep{2023MNRAS.520.3742K}. The high DM of FRBs, indicating their cosmological origins, can provide valuable information about the intergalactic medium (IGM) and the large-scale structure of the cosmos. Moreover, studying lensed FRBs, where the light from an FRB is bent due to the gravity of a massive object, like a black hole, along its path, can provide constraints on the fraction of primordial mass black holes comprised of dark matter \citep{2016PhRvL.117i1301M,2020ApJ...900..122S}.

\section{Effect of modified gravity in gravitational lensing}

In a strong lensing event, a single FRB source can be magnified and split into multiple images. Assuming spherical polar coordinates $(t,r,\theta,\phi)$, let us consider the following de Sitter–Schwarzschild metric 
\begin{align}
    \dd{s}^2 = \left(1-\frac{2GM_\mathrm{L}}{c^2r}+\kappa r^2\right)c^2\dd{t}^2 - \frac{1}{1-\frac{2GM_\mathrm{L}}{c^2r}+\kappa r^2}\dd{r}^2 -r^2 \dd\theta^2-r^2\sin^2\theta \dd\phi^2
\end{align}
with $M_\mathrm{L}$ being the lensed mass and $\kappa$ the modified gravity parameter. It is worth noting that this particular metric is a solution for a general scalar-tensor gravity theory framework. Thereby denoting $D_\mathrm{L}$ and $D_\mathrm{LS}$ respectively as the angular diameter distances from the observer to the lensing plane and from the lensing plane to the source, the differential time delay between the two lensed images is given by \cite{2023JCAP...11..059K} as
\begin{align}
    \Delta t = \left(1+z_\mathrm{L}\right) \left[\frac{1}{c\sqrt{\kappa}}\tan^{-1} \left(\sqrt{\kappa}\frac{4GM_\mathrm{L}}{c^2} \frac{y}{2}\sqrt{y^2+4}\right) + \frac{4GM_\mathrm{L}}{c^3} \ln\left(\frac{\sqrt{y^2+4} + y}{\sqrt{y^2+4} - y}\right)\right],
\end{align}
where $z_\mathrm{L}$ is the redshift of the lens plane, $y = r/\theta_\mathrm{E}\left(D_\mathrm{L}+D_\mathrm{LS}\right)$ with $\theta_\mathrm{E}$ being the Einstein radius.

In this analysis, we utilize 636 FRBs detected by the Canadian Hydrogen Intensity Mapping Experiment (CHIME) dataset\footnote{\url{https://www.chime-frb.ca/catalog}}. The total DM of each FRB can be decomposed into different components as follows:
\begin{align}
    \mathrm{DM} = \mathrm{DM}_\mathrm{MW} + \mathrm{DM}_\mathrm{Halo} + \mathrm{DM}_\mathrm{IGM}(z_\mathrm{S}) + \frac{\mathrm{DM}_\mathrm{Host}}{1+z_\mathrm{S}},
\end{align}
where $\mathrm{DM}_\mathrm{MW}$, $\mathrm{DM}_\mathrm{Halo}$, $\mathrm{DM}_\mathrm{IGM}$, and $\mathrm{DM}_\mathrm{Host}$ are respectively DM contributions from our Milky way galaxy, its circumgalactic halo, IGM, and host galaxy with $z_\mathrm{S}$ being the source redshift. Assuming standard $\Lambda$CDM cosmology, the mean $\mathrm{DM}_\mathrm{IGM}$ is given by \cite{2020Natur.581..391M} as
\begin{align}
    \langle \mathrm{DM}_\mathrm{IGM}\rangle = \frac{3cH_0^2\Omega_\mathrm{b}}{8\pi G m_\mathrm{p}} \int_{0}^{z_\mathrm{S}} \frac{f_\mathrm{IGM}(z)\chi(z)(1 + z)}{H(z)} \dd{z},
\end{align}
where $H(z) = H_0 \sqrt{\Omega_\mathrm{m} \left(1+z\right)^3 + \Omega_\Lambda}$ with $H_0$ being the Hubble constant, $\Omega_\mathrm{b}$ the baryonic matter density, $m_\mathrm{p}$ the proton mass, $\chi(z)$ the ionization fraction along the line of sight, and $f_\text{IGM}$ the baryon mass fraction in the IGM. Considering $\mathrm{DM}_\mathrm{MW}$ using the NE2001 model of Galactic distribution of free electrons, $\mathrm{DM}_\mathrm{Halo} \approx 50\rm\,pc\,cm^{-3}$, and $\langle\mathrm{DM}_\mathrm{Host}\rangle = 117\rm\,pc\,cm^{-3}$, we calculate $z_\mathrm{S}$ for all FRBs in the dataset using the aforementioned relation.

Taking into account the possibility of the lensed object being a primordial black hole (PBH), the optical depth of an FRB source is given by \cite{2016PhRvL.117i1301M} as
\begin{align}
    \tau(M_\mathrm{L},z_\mathrm{S}) &= \frac{3}{2c}f_\mathrm{PBH} \Omega_\mathrm{c}\int_0^{z_\mathrm{S}} \dd{z_\mathrm{L}} \frac{H_0^2}{H(z_\mathrm{L})} \frac{D_\mathrm{L} D_\mathrm{LS}}{D_\mathrm{S}} \left(1+z_\mathrm{L}\right)^2 \left[y_\mathrm{max}^2(\mu) - y_\mathrm{min}^2(M_\mathrm{L},z_\mathrm{L})\right],
\end{align}
with $f_\mathrm{PBH}$ being the fraction of dark matter made up of PBHs, $D_\mathrm{S}$ the angular diameter distance to the source, $\Omega_\mathrm{c}$ the current cold dark matter density, $\mu$ the magnification ratio, and $y_\mathrm{min}$ and $y_\mathrm{max}$ are the minimum and maximum impact parameters, respectively. Thereby, we define the integrated optical depth considering all FRBs as follows:
\begin{align}
    \bar{\tau} = \frac{1}{N_\mathrm{FRB}}\sum_{i=1}^{N_\mathrm{FRB}} \tau (M_\mathrm{L},z_{\mathrm{s},i}).
\end{align}
In general, the number of lensed FRBs is quite small in comparison to the total number of FRBs, $N_\mathrm{FRB}$. Thus utilizing Poisson statistics and taking into account for the fact that no lensed FRB has been confirmed so far, we obtain \citep{2023JCAP...11..059K}
\begin{align}
    f_\mathrm{PBH}<\frac{1}{\tau_1}\ln\left(\frac{N_\mathrm{FRB}}{N_\mathrm{FRB}-1}\right),
\end{align}
\begin{wrapfigure}{r}{0.5\textwidth}
\centering
\includegraphics[scale=0.4]{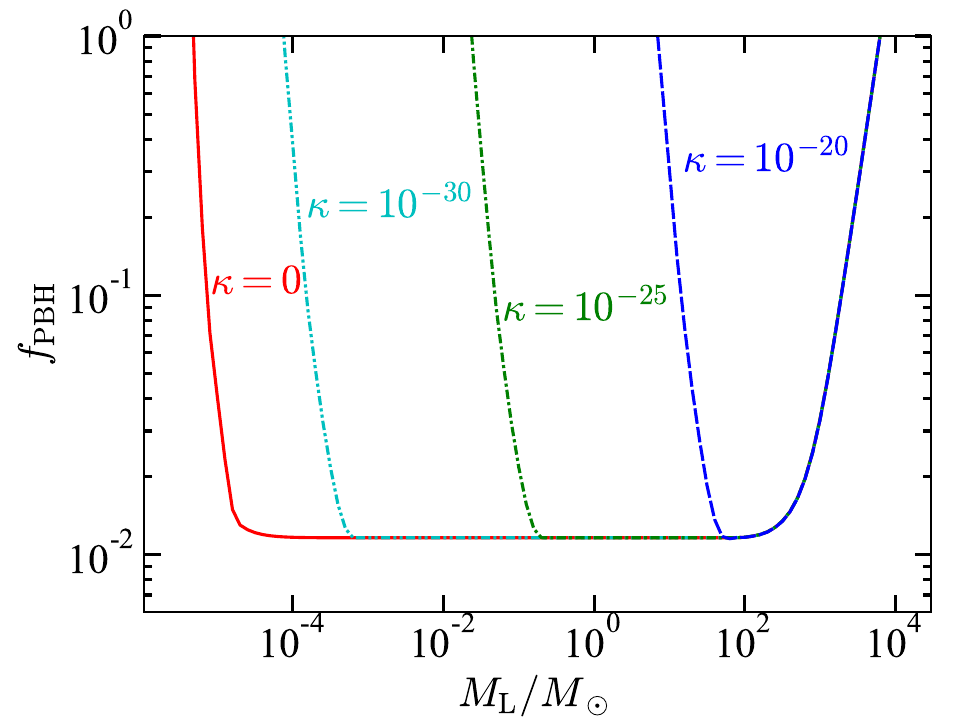}
\caption{Constraints on $f_\mathrm{PBH}$ for different values of $\kappa$ in cm$^{-2}$.}
\label{Fig:f_bound}
\end{wrapfigure}
where $\tau_1 = \bar{\tau}/f_\mathrm{PBH}$. Figure~\ref{Fig:f_bound} shows constraints on $f_\mathrm{PBH}$ for a diverse range of $\kappa$ values within the standard $\Lambda$CDM cosmological framework. It is interesting to note that the bound on $f_\mathrm{PBH}$ significantly depends on the value of the parameter $\kappa$. The minimum detectable time delay, approximately $10^{-9}\rm\,s$ for CHIME \citep{2022PhRvD.106d3017L}, gives the left-hand cutoff for each curve. On the other hand, the right-hand cutoff is determined by the maximum $\mu$ of each lensed image pair, which is $1<\mu<1/3\times \text{S/N}$ with $\text{S/N}$ being the signal-to-noise ratio. Interestingly, our results demonstrate similarity to those observed for plasma lensing by \cite{2022PhRvD.106d3017L}, which suggests that modified gravity mimics the behavior of a scattering screen of plasma lensing positioned along the path of the light ray.

\section{Conclusions}
This work explores how the gravitational lensing of FRBs can be utilized to constrain the fraction of PBHs composed of dark matter. It also confirms that modified gravity mimics the behavior of a scattering screen for light rays, similar to plasma lensing. In the future, next-generation radio telescopes with enhanced sensitivity and resolution, such as HIRAX, SKA, CHORD, DSA-2000, and BURSTT, hold immense promise for acquiring a wealth of new data. We anticipate not only improving the aforementioned constraint, but also leading to significant breakthroughs in FRB research.


\bibliographystyle{Bib}
\bibliography{Bibliography}

\end{document}